\documentclass[epjST_arXiv]{svjour_arXiv}

\usepackage[utf8]{inputenc}
\usepackage[english]{babel}
\usepackage{graphics}
\usepackage{color}
\usepackage{amssymb}
\usepackage{amsmath}
\usepackage{amsfonts}
\usepackage[numbers,sort&compress]{natbib}
\usepackage[colorlinks=true,linkcolor=blue,citecolor=blue,urlcolor=blue]{hyperref}

\begin{document}

\title{Quantum thermodynamics with a Josephson-photonics setup}

\author{Simon Dambach\hyperlink{test}{\footnote{\email{\href{mailto:simon.dambach@uni-ulm.de}{simon.dambach@uni-ulm.de}}\hypertarget{test}{}}} \and Paul Egetmeyer \and Joachim Ankerhold \and Björn Kubala}

\institute{Institute for Complex Quantum Systems and Center for Integrated Quantum Science and Technology, Ulm University, Albert-Einstein-Allee 11, 89081 Ulm, Germany}

\abstract{Josephson-photonics devices have emerged in the last years as versatile platforms to study light-charge interactions far from equilibrium and to create nonclassical radiation. Their potential to operate as nanoscale heat engines has also been discussed. The complementary mode of cooling is investigated here in the regimes of low and large photon occupancy, where nonlinearities are essential.
} 

\maketitle

\section{Introduction}
\label{sec_Introduction}
The experimental realization of thermal machines operating close to or even deep in the quantum regime is still in its infancy. Among others, particular progress has been made with superconducting platforms~\cite{Pekola2015}. Basic elements in these devices are superconducting tunnel junctions, implemented as Josephson junctions (JJ), with intrinsic nonlinearity and robustness against quasiparticle excitations. Recently, Josephson photonics has emerged as a fascinating new field to explore charge-light interactions far from equilibrium~\cite{Hofheinz2011,Chen2014,Cassidy2017}. There, the electrical energy carried by Cooper pairs tunneling through a dc-voltage-biased JJ is fully converted into photonic excitation quanta of microwave resonators. The details of this conversion and the feedback of the electromagnetic field onto the JJ can be controlled by the Josephson coupling energy and the impedance of the resonator; the latter one,  playing the role of an effective fine-structure constant, can by design even approach values of order unity. 
The situation where in presence of two detuned resonators a single Cooper pair creates simultaneously two photons, one in each of them, gives rise to nonclassical radiation~\cite{Westig2017}, but may also serve as a simple realization of a heat engine~\cite{Hofer2016a}, cf. also~\cite{Hofer2016b,Hofer2018,Leppaekangas2018}. Here, we briefly discuss the complementary mode, when it operates as a cooling device (power refrigerator). 
We address the conventional regime of low photon production as well as the semiclassical domain, where the performance of the machine is strongly influenced by nonlinear dynamics.

\section{Josephson-photonics setup as quantum refrigerator}
\label{sec_Section_Josephson-photonics_setup_as_quantum_refrigerator}

\begin{figure}
\centering
\resizebox{0.75\columnwidth}{!}{
\includegraphics{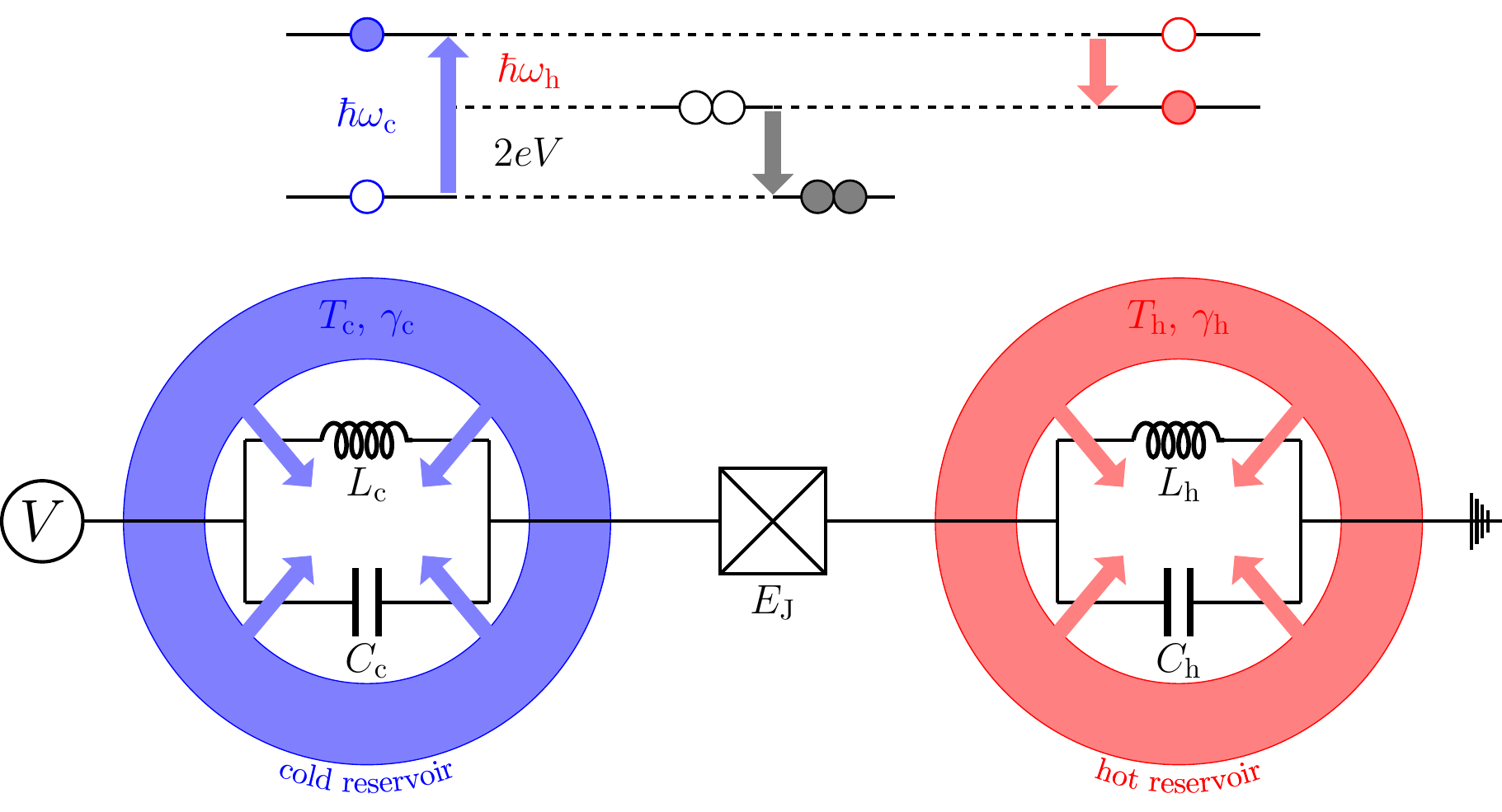}
}
\caption{Sketch of the circuit model and a schematic illustration of
the cooling process.}
\label{fig_1}
\end{figure}

We consider a system (see Fig.~\ref{fig_1}) consisting of a Josephson junction biased by an external dc voltage $V$ and two series-connected LC oscillators, denoted by c and h, with frequencies $\omega_{\mathrm{c(h)}}=1/\sqrt{L_{\mathrm{c(h)}}C_{\mathrm{c(h)}}}$. The two oscillators are coupled to thermal reservoirs with temperature $T_{\rm c}$ and $T_{\rm h}$, respectively, which are characterized by the occupation number $N_{\mathrm{c(h)}}=[\exp{(\hbar\omega_{\mathrm{c(h)}}/k_{\rm B}T_{\mathrm{c(h)}})}-1]^{-1}$.   

The setup can be modeled by the effective Hamiltonian~\cite{Armour2015,Trif2015}
\begin{equation}
H=\hbar\omega_{\rm c}n_{\rm c}+\hbar\omega_{\rm h}n_{\rm h}-E_{\rm J}\cos{\left(\omega_{\rm J}\tau+\phi_{\rm c}+\phi_{\rm h}\right)},
\end{equation}
where phase $\phi_{q}=\Delta_{q}(a_{q}^{\dagger}+a_{q})$ and photonic number operator $n_{q}=a_{q}^{\dagger}a_{q}$ of oscillator $q=\lbrace\mathrm{c},\mathrm{h}\rbrace$ are given in terms of the standard bosonic creation and annihilation operators $a^{\dagger}_{q}$ and $a_{q}$ with $[a_{q},a^{\dagger}_{q}]=1$. The parameter $\Delta_{q}=(2e^{2}Z_{q}/\hbar)^{1/2}$ gives the zero-point quantum fluctuations of the phase of oscillator $q$ with $Z_{q}=(L_{q}/C_{q})^{1/2}$ being the impedance.

By tuning the voltage $V$, we access the resonance $\omega_{\rm J}=2eV/\hbar\approx\omega_{\rm c}-\omega_{\rm h}$, where a Cooper-pair tunneling event across the junction is accompanied by the absorption of a photon from cavity $\mathrm{h}$ and the emission of a photon into cavity $\mathrm{c}$. 
Close to this resonance, a (time-independent) rotating-wave Hamiltonian can be derived:
\begin{equation}
H_{\mathrm{RWA}}=\hbar\delta_{\rm c}n_{\rm c}+\hbar\delta_{\rm h}n_{\rm h}
+\frac{E^{*}_{\rm J}}{2}:\left(a^{\dagger}_{\rm c}a_{\rm h}+a_{\rm c}a^{\dagger}_{\rm h}\right)\frac{J_{1}\!\left(2\Delta_{\rm c} \sqrt{n_{\rm c}}\right)J_{1}\!\left(2\Delta_{\rm h}\sqrt{n_{\rm h}}\right)}{\sqrt{n_{\rm c}}\sqrt{n_{\rm h}}}:
\label{eq_Hamiltonian_RWA}
\end{equation}
with a renormalized Josephson energy $E_{\rm J}^{*}=E_{\rm J}\exp{[-(\Delta^{2}_{\rm c}+\Delta^{2}_{\rm h})/2]}$ and normal-ordered Bessel functions $J_{1}$ of the first kind. Detunings, $\delta_{q}=\tilde{\omega}_{q}-\omega_{q}$, are subjected to the constraint $\omega_{\rm J}=\tilde{\omega}_{\rm c}-\tilde{\omega}_{\rm h}$.

Taking into account the coupling to thermal reservoirs as well as leakage of excited photons into the microwave output ports after a lifetime $1/\gamma_{\mathrm{c(h)}}$, the dynamics of the density operator $\rho$ is described by the master equation~\cite{Armour2015}
\begin{equation}
\begin{aligned}
\frac{\mathrm{d}\rho}{\mathrm{d}\tau}=&-\frac{i}{\hbar}\left[H_{\mathrm{RWA}},\rho\right]\\&+\!\!\!\sum_{q=\mathrm{c,h}}\!\frac{\gamma_{q}}{2}\!\left[\left(N_{q}\!+\!1\right)\!\left(2a_{q}\rho a_{q}^{\dagger}\!-\!a_{q}^{\dagger}a_{q}\rho\!-\!\rho a_{q}^{\dagger}a_{q}\right)\!+\!N_{q}\!\left(2a_{q}^{\dagger}\rho a_{q}\!-\!a_{q}a_{q}^{\dagger}\rho\!-\!\rho a_{q}a_{q}^{\dagger}\right)\right].
\label{eq_quantum_master_equation}
\end{aligned}
\end{equation}

Considering energy conservation immediately leads to the energy balance relation
\begin{equation}
\gamma_{\rm c}(\langle n_{\rm c}\rangle_{\mathrm{st}}-N_{\rm c})=\gamma_{\rm h}(N_{\rm h}-\langle n_{\rm h}\rangle_{\mathrm{st}})
\label{eq_energy_balance_relation}
\end{equation}
so that the steady-state mean photon occupations $\langle n_{\rm c}\rangle_{\mathrm{st}}$ and $\langle n_{\rm h}\rangle_{\mathrm{st}}$ are linked. 

\section{The low-photon limit}
\label{sec_The_low-photon_limit}

\begin{figure}
\centering
\resizebox{1.0\columnwidth}{!}{
\includegraphics{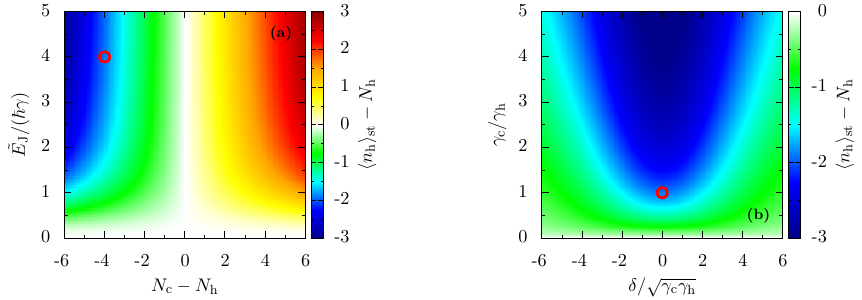}
}
\caption{Deviation of the mean steady-state photon occupation number $\langle n_{\rm h}\rangle_{\mathrm{st}}$ from thermal occupation $N_{\rm h}$ in the low-photon limit [Eq.~\eqref{eq_low-photon_result}]. Negative (positive) values indicate cooling (heating) of cavity $\mathrm{h}$. Plot (a) (with fixed parameters $\delta=\delta_{\rm c}-\delta_{\rm h}=0$ and $\gamma_{\rm c}=\gamma_{\rm h}=\gamma$) demonstrates that cooling of oscillator $\mathrm{h}$ necessarily requires $N_{\rm c}<N_{\rm h}$, i.e., $T_{\rm c}/\omega_{\rm c}<T_{\rm h}/\omega_{\rm h}$~\cite{Hofer2016a}.  The effect of cooling is enhanced with increasing driving $E_{\rm J}$; in the limit of strong driving, $\tilde{E}_{\rm}/(\hbar\gamma)\rightarrow\infty$, the photon occupation of the two cavities equilibrates: $\langle n_{\rm c}\rangle_{\mathrm{st}}=\langle n_{\rm h}\rangle_{\mathrm{st}}=(N_{\rm c}+N_{\rm h})/2$. Plot (b) illustrates the influence of finite detuning, $\delta\neq0$, and asymmetric decay rates, $\gamma_{\rm c}/\gamma_{\rm h}\neq1$, on the cooling effect in cavity $\mathrm{h}$ for parameters $\tilde{E}_{\rm J}/(\hbar\sqrt{\gamma_{\rm c}\gamma_{\rm h}})=4$ and $N_{\rm c}-N_{\rm h}=-4$ (red circle). While detuning reduces the cooling, an increased coupling $\gamma_{\rm c}>\gamma_{\rm h}$ leads to stronger cooling since cavity $\mathrm{c}$ stays closer to the thermal equilibrium with the cold reservoir.}
\label{fig_2}
\end{figure}

In the regime of sufficiently low photon occupation where $2\Delta_{\rm c(h)}\sqrt{\langle n_{\mathrm{c(h)}}\rangle}\ll1$, we can expand the Bessel functions in the Hamiltonian [Eq.~\eqref{eq_Hamiltonian_RWA}] to lowest order: 
\begin{equation}
H^{(0)}_{\mathrm{RWA}}=\hbar\delta_{\rm c}n_{\rm c}+\hbar\delta_{\rm h}n_{\rm h}
+\frac{\tilde{E}_{\rm J}}{2}\left(a^{\dagger}_{\rm c}a_{\rm h}+a_{\rm c}a^{\dagger}_{\rm h}\right)
\end{equation}
with $\tilde{E}_{\rm J}=\Delta_{\rm c}\Delta_{\rm h}E^{*}_{J}$. Based on this linearized Hamiltonian $H^{(0)}_{\mathrm{RWA}}$ and the quantum master equation in Eq.~\eqref{eq_quantum_master_equation}, equations of motions are derived which yield the steady-state mean occupations of the cavities 
\begin{equation}
\langle n_{\rm h}\rangle_{\mathrm{st}}-N_{\rm h}=\frac{(N_{\rm c}-N_{\rm h})(1+\gamma_{\rm c}/\gamma_{\rm h})}{(2+\gamma_{\rm c}/\gamma_{\rm h}+\gamma_{\rm h}/\gamma_{\rm c})[1+\hbar^{2}\gamma_{\rm c}\gamma_{\rm h}/\tilde{E}_{\rm J}^{2}]+4[(\delta_{\rm c}-\delta_{\rm h})\hbar\sqrt{\gamma_{\rm c}\gamma_{\rm h}}/\tilde{E}_{\rm J}]^{2}},\label{eq_low-photon_result}
\end{equation}
where $\langle n_{\rm c}\rangle_{\mathrm{st}}$ follows from $\langle n_{\rm h}\rangle_{\mathrm{st}}$ by exchanging $\mathrm{c}$ and $\mathrm{h}$. Figure~\ref{fig_2} illustrates various cooling and heating scenarios based on this analytical result.

\section{The semiclassical limit}
\label{sec_The_semiclassical_limit}

\begin{figure}
\centering
\resizebox{1.0\columnwidth}{!}{
\includegraphics{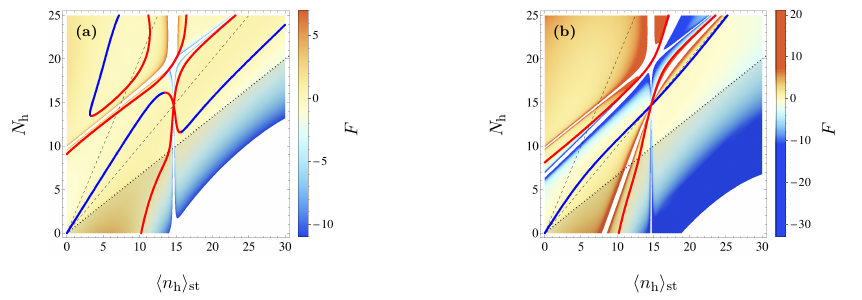}
}
\caption{Self-consistency function $F$ and stable (blue) and unstable solutions (red) [$F=0$ $\Leftrightarrow$ $\partial_{\tau}(\langle n_{\rm c}\rangle,\,\langle n_{\rm h}\rangle,\,\langle a_{\rm c}^{\dagger}a_{\rm h}\rangle)^{T}=0$] (a) without and (b) with detuning $\delta=0.16E_{\rm J}/\hbar$; other parameters are $\gamma_{\rm c}=\gamma_{\rm h}=0.12E_{\rm J}/\hbar$ and $\Delta_{\rm c}=\Delta_{\rm h}=0.5$ and $N_{\rm c}=N_{\rm h}/2$.
Dashed lines indicate $\langle n_{\rm h}\rangle_{\mathrm{st}}=N_{\rm h}$ and $\langle n_{\rm h}\rangle_{\mathrm{st}}=N_{\rm c}=N_{\rm h}/2$. For values $\langle n_{\rm h}\rangle_{\mathrm{st}}>\langle n_{\rm h}\rangle_{\mathrm{st}}^{\mathrm{max}}$ (shown as darker region), the solution for $\langle n_{\rm c}\rangle_{\mathrm{st}}$ is nonphysical (negative).
The structure of $F$ is determined by the vanishing of the transfer rate between the cavities [i.e., the second term in Eq.~\eqref{eq_semiclassical_n_c}] at zeros of the Bessel functions.
Such zeros lead to poles in $F$ (white regions) for certain fixed values of $\langle n_{\rm h}\rangle_{\mathrm{st}}\approx14.68$ (vertical white regions) or if $\langle n_{\rm c}\rangle_{\mathrm{st}}$ takes the same value [sloped white region, cf.~Eq.~\eqref{eq_energy_balance_relation}]. For finite detuning, additional poles appear.
Interestingly,  multiple stable solutions for $\langle n_{\rm h}\rangle_{\mathrm{st}}$ exist for certain fixed parameters $N_{\rm h}$. Such solutions are separated by a crossing of a Bessel zero and correspondingly a sign change of the transition amplitude in the equation of motion [Eq.~\eqref{eq_semiclassical_n_c}].
}    
\label{fig_3}
\end{figure}

\begin{figure}
\centering
\resizebox{0.65\columnwidth}{!}{
\includegraphics{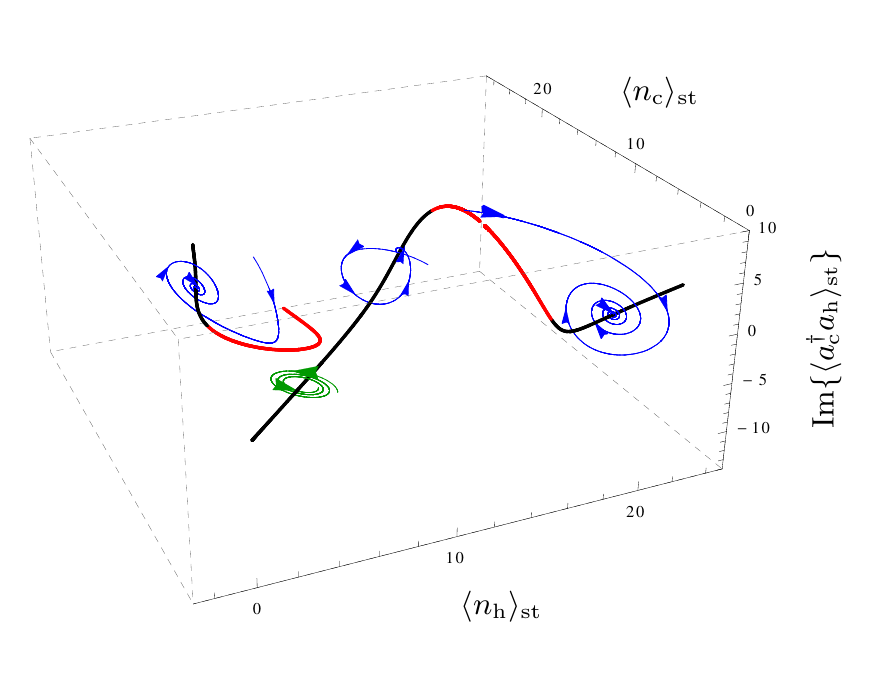}
}
\caption{
Semiclassical trajectories of the system state $(\langle n_{\rm c}\rangle,\,\langle n_{\rm h}\rangle,\,\langle a_{\rm c}^\dagger a_{\rm h}\rangle)^{T}$ for fixed values of the system parameters: $\Delta_{\rm c}=\Delta_{\rm h}=0.5$, $\gamma_{\rm c}=\gamma_{\rm h}=0.12E_{\rm J}/\hbar$, $\delta=0$ so that $\mathrm{Re}\lbrace\langle a_{\rm c}^\dagger a_{\rm h}\rangle\rbrace=0$, $N_{\rm h}=2N_{\rm c}=15.5$ (blue) and $5$ (green). Thick lines indicate the stable (black) and unstable (red) solutions for $N_{\rm h}=2N_c \in [2,25]$, also visible in Fig.~\hyperref[fig_3]{3(a)} (cf. the upper left and the S-shaped curve in that plot). Depending on the initial system state, different stable solutions are reached. In the low-photon regime, $N_{\rm h}=5$  (green), starting from an initial thermal occupation of the cavities (e.g., by $E_{\rm J}(\tau<0)=0$) always the trivial steady state with $N_{\rm c}<\langle n_{\rm c}\rangle_{\mathrm{st}}<\langle n_{\rm h}\rangle_{\mathrm{st}}<N_{\rm h}$ is reached. To reach a desired stationary state (e.g., the `ultra-cold' solution with $\langle n_{\rm h}\rangle_{\mathrm{st}}<N_{\rm c}$ from an initial thermal state), parameter control, which is more complex than a sudden switching on of $E_{\rm J}$ at $\tau=0$, may be employed.    
}
\label{fig_4}
\end{figure}

To explore the regime of higher photon numbers, we turn toward a semiclassical description, i.e., we assume that quantum fluctuations $\Delta_{\mathrm{c(h)}}$ are small. Assuming a Gaussian Wigner density for the cavity states, equations of motions for the second moments,  
\begin{equation}
\frac{\mathrm{d}\langle n_{\rm c}\rangle}{\mathrm{d}\tau}\!=\!\gamma_{\rm c}(\!N_{\rm c}-\langle n_{\rm c}\rangle\!)+\frac{E^{*}_{\rm J}}{\hbar}\mathrm{Im}\lbrace\langle a_{\rm c}^{\dagger}a_{\rm h}\rangle\rbrace\frac{\!J_{1}\!\left(\!2\Delta_{\rm c} \sqrt{\!\langle n_{\rm c}\rangle\!}\right)\!J_{1}\!\left(\!2\Delta_{\rm h}\sqrt{\!\langle n_{\rm h}\rangle\!}\right)\!}{\sqrt{\!\langle n_{\rm c}\rangle\!}\sqrt{\!\langle n_{\rm h}\rangle\!}}\quad\!\!\!(\textrm{and c}\leftrightarrow\mathrm{h})\label{eq_semiclassical_n_c},
\end{equation}
\begin{align}
\frac{\mathrm{d}\langle a^{\dagger}_{\rm c}a_{\rm h}\rangle}{\mathrm{d}\tau}&=\left[i\delta-\frac{1}{2}(\gamma_{\rm c}+\gamma_{\rm h})\right]\langle a^{\dagger}_{\rm c}a_{\rm h}\rangle\nonumber\\
&-i\frac{E^{*}_{\rm J}}{2\hbar}\Bigg[\langle a_{\rm c}^{\dagger}a_{\rm h}\rangle^{2}\left(\Delta_{\rm c}\frac{J_{2}\!\left(\!2\Delta_{\rm c} \sqrt{\!\langle n_{\rm c}\rangle\!}\right)J_{1}\!\left(\!2\Delta_{\rm h}\sqrt{\!\langle n_{\rm h}\rangle\!}\right)}{\langle n_{\rm c}\rangle\sqrt{\!\langle n_{\rm h}\rangle\!}}-\mathrm{c}\leftrightarrow\mathrm{h}\right)\\
&\phantom{\,i\frac{E^{*}_{\rm J}}{2\hbar}\Bigg[\langle a_{\rm c}^{\dagger}a_{\rm h}\rangle^{2}}-\left(\Delta_{\rm c}\langle n_{\rm h}\rangle\frac{J_{0}\!\left(\!2\Delta_{\rm c} \sqrt{\!\langle n_{\rm c}\rangle\!}\right)J_{1}\!\left(\!2\Delta_{\rm h}\sqrt{\!\langle n_{\rm h}\rangle\!}\right)}{\sqrt{\!\langle n_{\rm h}\rangle\!}}-\mathrm{c}\leftrightarrow\mathrm{h}\right)\Bigg],\nonumber
\end{align}
result. Stationary solutions for the cavity state $(\langle n_{\rm c}\rangle,\,\langle n_{\rm h}\rangle,\,\langle a_{\rm c}^{\dagger}a_{\rm h}\rangle)^{T}$ are found from these equations by iteratively eliminating the other variables so that finally a self-consistent equation $F(\langle n_{\rm h}\rangle_{\mathrm{st}})=0$ for $\langle n_{\rm h}\rangle_{\mathrm{st}}$ alone is obtained. Note that from the energy balance equation [Eq.~\eqref{eq_energy_balance_relation}] a limitation $\langle n_{\rm h}\rangle_{\mathrm{st}}<\langle n_{\rm h}\rangle_{\mathrm{st}}^{\mathrm{max}}=\gamma_{\rm c}N_{\rm c}+\gamma_{\rm h}N_{\rm h}$ results from requiring a positive $\langle n_{\rm c}\rangle_{\mathrm{st}}$. The self-consistency function $F$ can be employed to understand the structure of steady-state solutions, which becomes complicated (e.g., multivalued) beyond the linear regime as shown for various parameters in Fig.~\ref{fig_3}. Stability and relaxation toward the steady state is further discussed in Fig.~\ref{fig_4}.         

\section{Conclusions}
\label{sec_Conclusions}
We investigated the possibility of using a Josephson-photonics setup, where two microwave-stripline cavities are connected by a dc-biased junction, to cool down one of the cavities.
In a process similar to sideband cooling, an excitation from the cooling target is combined with the energy provided by the bias to deposit a photon in the heat dump.

In the low-photon regime, the cavity with higher occupation will be cooled down as the mean occupations of the two cavities are pulled closer by increasing Josephson coupling.
For higher occupations, the full nonlinearity of the Josephson junction comes into play and our semiclassical analysis predicts a novel `ultra-cold' solution. Stability and accessibility of this solution in the full quantum regime as well as the effect of multistability on the quantum statistics of the heat transfer remain to be explored. Josephson-photonics setups may offer unique insight into the statistics of quantum heat flow due to the possibility of observing photon emission to the reservoirs.
\begin{acknowledgement}
The authors would like to thank M.I. Dykman
and F. Portier for valuable discussions. This work was supported by the Center for Integrated Quantum Science and Technology ($\mathrm{IQ}^\mathrm{ST}$) and by the Deutsche Forschungsgemeinschaft (DFG) through Grant No. AN336/11-1. S.D. acknowledges financial support from the Carl-Zeiss-Stiftung.\newline

\noindent
All authors participated in the planning of the project and were involved in the analysis and interpretation of the results. S.D., J.A., and B.K. wrote the manuscript.
\end{acknowledgement}

\end{document}